# 1. THE COSMOLOGICAL PARAMETERS

Updated September 2009, by O. Lahav (University College London) and A.R. Liddle (University of Sussex).

## 1.1. Parametrizing the Universe

Rapid advances in observational cosmology have led to the establishment of a precision cosmological model, with many of the key cosmological parameters determined to one or two significant figure accuracy. Particularly prominent are measurements of cosmic microwave background (CMB) anisotropies, led by the five-year results from the Wilkinson Microwave Anisotropy Probe (WMAP) [1,2,3]. However the most accurate model of the Universe requires consideration of a wide range of different types of observation, with complementary probes providing consistency checks, lifting parameter degeneracies, and enabling the strongest constraints to be placed.

The term 'cosmological parameters' is forever increasing in its scope, and nowadays includes the parametrization of some functions, as well as simple numbers describing properties of the Universe. The original usage referred to the parameters describing the global dynamics of the Universe, such as its expansion rate and curvature. Also now of great interest is how the matter budget of the Universe is built up from its constituents: baryons, photons, neutrinos, dark matter, and dark energy. We need to describe the nature of perturbations in the Universe, through global statistical descriptors such as the matter and radiation power spectra. There may also be parameters describing the physical state of the Universe, such as the ionization fraction as a function of time during the era since recombination. Typical comparisons of cosmological models with observational data now feature between five and ten parameters.

### 1.1.1. *The global description of the Universe*:

Ordinarily, the Universe is taken to be a perturbed Robertson–Walker space-time with dynamics governed by Einstein's equations. This is described in detail by Olive and Peacock in this volume. Using the density parameters $\Omega_i$ for the various matter species and $\Omega_\Lambda$ for the cosmological constant, the Friedmann equation can be written

$$\sum_i \Omega_i + \Omega_\Lambda - 1 = \frac{k}{R^2 H^2}, \qquad (1.1)$$

where the sum is over all the different species of material in the Universe. This equation applies at any epoch, but later in this article we will use the symbols $\Omega_i$ and $\Omega_\Lambda$ to refer to the present values. A typical collection would be baryons, photons, neutrinos, and dark matter (given charge neutrality, the electron density is guaranteed to be too small to be worth considering separately and is included with the baryons).

The complete present state of the homogeneous Universe can be described by giving the current values of all the density parameters and of the Hubble parameter $h$. These also allow us to track the history of the Universe back in time, at least until an epoch where interactions allow interchanges between the densities of the different species, which is believed to have last happened at neutrino decoupling, shortly before Big Bang Nucleosynthesis (BBN). To probe further back into the Universe's history requires assumptions about particle interactions, and perhaps about the nature of physical laws themselves.

### 1.1.2. *Neutrinos*:

The standard neutrino sector has three flavors. For neutrinos of mass in the range $5 \times 10^{-4}$ eV to 1 MeV, the density parameter in neutrinos is predicted to be

$$\Omega_\nu h^2 = \frac{\sum m_\nu}{93\,\mathrm{eV}}, \tag{1.2}$$

where the sum is over all families with mass in that range (higher masses need a more sophisticated calculation). We use units with $c = 1$ throughout. Results on atmospheric and Solar neutrino oscillations [4] imply non-zero mass-squared differences between the three neutrino flavors. These oscillation experiments cannot tell us the absolute neutrino masses, but within the simple assumption of a mass hierarchy suggest a lower limit of $\Omega_\nu \approx 0.001$ on the neutrino mass density parameter.

For a total mass as small as 0.1 eV, this could have a potentially observable effect on the formation of structure, as neutrino free-streaming damps the growth of perturbations. Present cosmological observations have shown no convincing evidence of any effects from either neutrino masses or an otherwise non-standard neutrino sector, and impose quite stringent limits, which we summarize in Section 1.3.4. Accordingly, the usual assumption is that the masses are too small to have a significant cosmological impact at present data accuracy. However, we note that the inclusion of neutrino mass as a free parameter can affect the derived values of other cosmological parameters.

The cosmological effect of neutrinos can also be modified if the neutrinos have decay channels, or if there is a large asymmetry in the lepton sector manifested as a different number density of neutrinos versus anti-neutrinos. This latter effect would need to be of order unity to be significant (rather than the $10^{-9}$ seen in the baryon sector), which may be in conflict with nucleosynthesis [5].

### 1.1.3. *Inflation and perturbations*:

A complete model of the Universe should include a description of deviations from homogeneity, at least in a statistical way. Indeed, some of the most powerful probes of the parameters described above come from the evolution of perturbations, so their study is naturally intertwined in the determination of cosmological parameters.

There are many different notations used to describe the perturbations, both in terms of the quantity used to describe the perturbations and the definition of the statistical measure. We use the dimensionless power spectrum $\Delta^2$ as defined in Olive and Peacock (also denoted $\mathcal{P}$ in some of the literature). If the perturbations obey Gaussian statistics, the power spectrum provides a complete description of their properties.

From a theoretical perspective, a useful quantity to describe the perturbations is the curvature perturbation $\mathcal{R}$, which measures the spatial curvature of a comoving slicing of the space-time. A case of particular interest is the Harrison–Zel'dovich spectrum, which corresponds to a constant $\Delta^2_\mathcal{R}$. More generally, one can approximate the spectrum by a power-law, writing

$$\Delta^2_\mathcal{R}(k) = \Delta^2_\mathcal{R}(k_*) \left[\frac{k}{k_*}\right]^{n-1}, \tag{1.3}$$

where $n$ is known as the spectral index, always defined so that $n = 1$ for the Harrison–Zel'dovich spectrum, and $k_*$ is an arbitrarily chosen scale. The initial spectrum, defined at some early epoch of the Universe's history, is usually taken to have a simple form such as this power-law, and we will see that observations require $n$ close to one, which corresponds to the perturbations in the curvature being independent of scale. Subsequent evolution will modify the spectrum from its initial form.

The simplest viable mechanism for generating the observed perturbations is the inflationary cosmology, which posits a period of accelerated expansion in the Universe's early stages [6]. It is a useful working hypothesis that this is the sole mechanism for generating perturbations, and it may further be assumed to be the simplest class of inflationary model, where the dynamics are equivalent to that of a single scalar field $\phi$ slowly rolling on a potential $V(\phi)$. One may seek to verify that this simple picture can match observations and to determine the properties of $V(\phi)$ from the observational data. Alternatively, more complicated models, perhaps motivated by contemporary fundamental physics ideas, may be tested on a model-by-model basis.

Inflation generates perturbations through the amplification of quantum fluctuations, which are stretched to astrophysical scales by the rapid expansion. The simplest models generate two types, density perturbations which come from fluctuations in the scalar field and its corresponding scalar metric perturbation, and gravitational waves which are tensor metric fluctuations. The former experience gravitational instability and lead to structure formation, while the latter can influence the CMB anisotropies. Defining slow-roll parameters, with primes indicating derivatives with respect to the scalar field, as

$$\epsilon = \frac{m_{\rm Pl}^2}{16\pi}\left(\frac{V'}{V}\right)^2 \quad ; \quad \eta = \frac{m_{\rm Pl}^2}{8\pi}\frac{V''}{V}, \tag{1.4}$$

which should satisfy $\epsilon, |\eta| \ll 1$, the spectra can be computed using the slow-roll approximation as

$$\begin{aligned}\Delta_{\mathcal{R}}^2(k) &\simeq \left.\frac{8}{3m_{\rm Pl}^4}\frac{V}{\epsilon}\right|_{k=aH} ; \\ \Delta_{\rm grav}^2(k) &\simeq \left.\frac{128}{3m_{\rm Pl}^4}V\right|_{k=aH}.\end{aligned} \tag{1.5}$$

In each case, the expressions on the right-hand side are to be evaluated when the scale $k$ is equal to the Hubble radius during inflation. The symbol '$\simeq$' here indicates use of the slow-roll approximation, which is expected to be accurate to a few percent or better.

From these expressions, we can compute the spectral indices

$$n \simeq 1 - 6\epsilon + 2\eta \quad ; \quad n_{\rm grav} \simeq -2\epsilon. \tag{1.6}$$

Another useful quantity is the ratio of the two spectra, defined by

$$r \equiv \frac{\Delta_{\rm grav}^2(k_*)}{\Delta_{\mathcal{R}}^2(k_*)}. \tag{1.7}$$

This convention matches that of recent versions of the CMBFAST code [7] and that used by WMAP [8] (there are some alternative historical definitions which lead to a slightly different prefactor in the following equation). We have

$$r \simeq 16\epsilon \simeq -8n_{\text{grav}}, \tag{1.8}$$

which is known as the consistency equation.

In general, one could consider corrections to the power-law approximation, which we discuss later. However, for now we make the working assumption that the spectra can be approximated by power laws. The consistency equation shows that $r$ and $n_{\text{grav}}$ are not independent parameters, and so the simplest inflation models give initial conditions described by three parameters, usually taken as $\Delta_\mathcal{R}^2$, $n$, and $r$, all to be evaluated at some scale $k_*$, usually the 'statistical center' of the range explored by the data. Alternatively, one could use the parametrization $V$, $\epsilon$, and $\eta$, all evaluated at a point on the putative inflationary potential.

After the perturbations are created in the early Universe, they undergo a complex evolution up until the time they are observed in the present Universe. While the perturbations are small, this can be accurately followed using a linear theory numerical code such as CMBFAST [7]. This works right up to the present for the CMB, but for density perturbations on small scales non-linear evolution is important and can be addressed by a variety of semi-analytical and numerical techniques. However the analysis is made, the outcome of the evolution is in principle determined by the cosmological model, and by the parameters describing the initial perturbations, and hence can be used to determine them.

Of particular interest are CMB anisotropies. Both the total intensity and two independent polarization modes are predicted to have anisotropies. These can be described by the radiation angular power spectra $C_\ell$ as defined in the article of Scott and Smoot in this volume, and again provide a complete description if the density perturbations are Gaussian.

**1.1.4.   *The standard cosmological model*:**

We now have most of the ingredients in place to describe the cosmological model. Beyond those of the previous subsections, there are two parameters which are essential — a measure of the ionization state of the Universe and the galaxy bias parameter. The Universe is known to be highly ionized at low redshifts (otherwise radiation from distant quasars would be heavily absorbed in the ultra-violet), and the ionized electrons can scatter microwave photons altering the pattern of observed anisotropies. The most convenient parameter to describe this is the optical depth to scattering $\tau$ (*i.e.*, the probability that a given photon scatters once); in the approximation of instantaneous and complete reionization, this could equivalently be described by the redshift of reionization $z_{\text{ion}}$. The bias parameter, described fully later, is needed to relate the observed galaxy power spectrum to the predicted dark matter power spectrum.    The basic set of cosmological parameters is therefore as shown in Table 1.1. The spatial curvature does not appear in the list, because it can be determined from the other parameters using

**Table 1.1:** The basic set of cosmological parameters. We give values (with some additional rounding) as obtained using a fit of a ΛCDM cosmology with a power-law initial spectrum to WMAP5 data alone [2]. Tensors are assumed zero except in quoting a limit on them. The exact values and uncertainties depend on both the precise data-sets used and the choice of parameters allowed to vary (see Table 1.2 for the former). Limits on $\Omega_\Lambda$ and $h$ weaken if the Universe is not assumed flat. The density perturbation amplitude is specified by the derived parameter $\sigma_8$. Uncertainties are one-sigma/68% confidence unless otherwise stated.

| Parameter | Symbol | Value |
|---|---|---|
| Hubble parameter | $h$ | $0.72 \pm 0.03$ |
| Total matter density | $\Omega_m$ | $\Omega_m h^2 = 0.133 \pm 0.006$ |
| Baryon density | $\Omega_b$ | $\Omega_b h^2 = 0.0227 \pm 0.0006$ |
| Cosmological constant | $\Omega_\Lambda$ | $\Omega_\Lambda = 0.74 \pm 0.03$ |
| Radiation density | $\Omega_r$ | $\Omega_r h^2 = 2.47 \times 10^{-5}$ |
| Neutrino density | $\Omega_\nu$ | See Sec. 1.1.2 |
| Density perturbation amplitude | $\Delta_\mathcal{R}^2(k = 0.002\,\text{Mpc})$ | $(2.41 \pm 0.11) \times 10^{-9}$ |
| Density perturbation spectral index | $n$ | $n = 0.963^{+0.014}_{-0.015}$ |
| Tensor to scalar ratio | $r$ | $r < 0.43$ (95% conf.) |
| Ionization optical depth | $\tau$ | $\tau = 0.087 \pm 0.017$ |
| Bias parameter | $b$ | See Sec. 1.3.4 |

Eq. (1.1). The total present matter density $\Omega_m = \Omega_{cdm} + \Omega_b$ is usually used in place of the dark matter density.

Most attention to date has been on parameter estimation, where a set of parameters is chosen by hand and the aim is to constrain them. Interest has been growing towards the higher-level inference problem of model selection, which compares different choices of parameter sets. Bayesian inference offers an attractive framework for cosmological model selection, setting a tension between model predictiveness and ability to fit the data.

As described in Sec. 1.4, models based on these eleven parameters are able to give a good fit to the complete set of high-quality data available at present, and indeed some simplification is possible. Observations are consistent with spatial flatness, and indeed the inflation models so far described automatically generate negligible spatial curvature, so we can set $k = 0$; the density parameters then must sum to unity, and so one can be eliminated. The neutrino energy density is often not taken as an independent parameter. Provided the neutrino sector has the standard interactions, the neutrino energy density, while relativistic, can be related to the photon density using thermal physics arguments, and it is currently difficult to see the effect of the neutrino mass, although observations of large-scale structure have already placed interesting upper limits. This reduces the standard parameter set to nine. In addition, there is no observational evidence for the existence of tensor perturbations (though the upper limits are quite weak), and so $r$ could be set to zero. Presently $n$ is in a somewhat controversial position regarding whether it needs to be varied in a fit, or can be set to the Harrison–Zel'dovich value $n = 1$. Parameter estimation [2] suggests $n = 1$ is ruled out at some significance, but Bayesian model selection techniques [9] suggest the data is not conclusive. With $n$ set to one, this leaves seven parameters, which is the smallest set that can usefully be compared to the present cosmological data set. This model (usually with $n$ kept as a parameter) is referred to by various names, including $\Lambda$CDM, the concordance cosmology, and the standard cosmological model.

Of these parameters, only $\Omega_r$ is accurately measured directly. The radiation density is dominated by the energy in the CMB, and the COBE satellite FIRAS experiment determined its temperature to be $T = 2.725 \pm 0.001$ K [10], corresponding to $\Omega_r = 2.47 \times 10^{-5} h^{-2}$. It typically need not be varied in fitting other data. If galaxy clustering data are not included in a fit, then the bias parameter is also unnecessary.

In addition to this minimal set, there is a range of other parameters which might prove important in future as the data-sets further improve, but for which there is so far no direct evidence, allowing them to be set to a specific value for now. We discuss various speculative options in the next section. For completeness at this point, we mention one other interesting parameter, the helium fraction, which is a non-zero parameter that can affect the CMB anisotropies at a subtle level. Presently, BBN provides the best measurement of this parameter (see the Fields and Sarkar article in this volume), and it is usually fixed in microwave anisotropy studies, but the data are just reaching a level where allowing its variation may become mandatory.

**1.1.5.  *Derived parameters*:**

The parameter list of the previous subsection is sufficient to give a complete description of cosmological models which agree with observational data. However, it is not a unique parametrization, and one could instead use parameters derived from that basic set. Parameters which can be obtained from the set given above include the age of the Universe, the present horizon distance, the present neutrino background temperature, the epoch of matter–radiation equality, the epochs of recombination and decoupling, the epoch of transition to an accelerating Universe, the baryon-to-photon ratio, and the baryon to dark matter density ratio. In addition, the physical densities of the matter components, $\Omega_i h^2$, are often more useful than the density parameters. The density perturbation amplitude can be specified in many different ways other than the large-scale primordial amplitude, for instance, in terms of its effect on the CMB, or by specifying a short-scale quantity, a common choice being the present linear-theory mass dispersion on a scale of $8\,h^{-1}$Mpc, known as $\sigma_8$, whose WMAP5 value is $0.80 \pm 0.04$.

Different types of observation are sensitive to different subsets of the full cosmological parameter set, and some are more naturally interpreted in terms of some of the derived parameters of this subsection than on the original base parameter set. In particular, most types of observation feature degeneracies whereby they are unable to separate the effects of simultaneously varying several of the base parameters.

## 1.2. Extensions to the standard model

This section discusses some ways in which the standard model could be extended. At present, there is no positive evidence in favor of any of these possibilities, which are becoming increasingly constrained by the data, though there always remains the possibility of trace effects at a level below present observational capability.

### 1.2.1. *More general perturbations*:

The standard cosmology assumes adiabatic, Gaussian perturbations. Adiabaticity means that all types of material in the Universe share a common perturbation, so that if the space-time is foliated by constant-density hypersurfaces, then all fluids and fields are homogeneous on those slices, with the perturbations completely described by the variation of the spatial curvature of the slices. Gaussianity means that the initial perturbations obey Gaussian statistics, with the amplitudes of waves of different wavenumbers being randomly drawn from a Gaussian distribution of width given by the power spectrum. Note that gravitational instability generates non-Gaussianity; in this context, Gaussianity refers to a property of the initial perturbations, before they evolve significantly.

The simplest inflation models, based on one dynamical field, predict adiabatic fluctuations and a level of non-Gaussianity which is too small to be detected by any experiment so far conceived. For present data, the primordial spectra are usually assumed to be power laws.

#### 1.2.1.1. *Non-power-law spectra:*

For typical inflation models, it is an approximation to take the spectra as power laws, albeit usually a good one. As data quality improves, one might expect this approximation to come under pressure, requiring a more accurate description of the initial spectra, particularly for the density perturbations. In general, one can write a Taylor expansion of $\ln \Delta_{\mathcal{R}}^2$ as

$$\ln \Delta_{\mathcal{R}}^2 (k) = \ln \Delta_{\mathcal{R}}^2 (k_*) + (n_* - 1) \ln \frac{k}{k_*} + \frac{1}{2} \left. \frac{dn}{d \ln k} \right|_* \ln^2 \frac{k}{k_*} + \cdots, \tag{1.9}$$

where the coefficients are all evaluated at some scale $k_*$. The term $dn/d\ln k|_*$ is often called the running of the spectral index [11]. Once non-power-law spectra are allowed, it is necessary to specify the scale $k_*$ at which the spectral index is defined.

#### 1.2.1.2. *Isocurvature perturbations:*

An isocurvature perturbation is one which leaves the total density unperturbed, while perturbing the relative amounts of different materials. If the Universe contains $N$ fluids, there is one growing adiabatic mode and $N-1$ growing isocurvature modes (for reviews see Ref. 12 and Ref. 13). These can be excited, for example, in inflationary models where there are two or more fields which acquire dynamically-important perturbations. If one field decays to form normal matter, while the second survives to become the dark matter, this will generate a cold dark matter isocurvature perturbation.

In general, there are also correlations between the different modes, and so the full set of perturbations is described by a matrix giving the spectra and their correlations.

Constraining such a general construct is challenging, though constraints on individual modes are beginning to become meaningful, with no evidence that any other than the adiabatic mode must be non-zero.

**1.2.1.3.** *Non-Gaussianity:*

Multi-field inflation models can also generate primordial non-Gaussianity (reviewed, *e.g.*, in Ref. 13). The extra fields can either be in the same sector of the underlying theory as the inflaton, or completely separate, an interesting example of the latter being the curvaton model [14]. Current upper limits on non-Gaussianity are becoming stringent, but there remains much scope to push down those limits and perhaps reveal trace non-Gaussianity in the data. If non-Gaussianity is observed, its nature may favor an inflationary origin, or a different one such as topological defects. A plausible possibility is non-Gaussianity caused by defects forming in a phase transition which ended inflation.

**1.2.2.** *Dark matter properties*:

Dark matter properties are discussed in the article by Drees and Gerbier in this volume. The simplest assumption concerning the dark matter is that it has no significant interactions with other matter, and that its particles have a negligible velocity as far as structure formation is concerned. Such dark matter is described as 'cold,' and candidates include the lightest supersymmetric particle, the axion, and primordial black holes. As far as astrophysicists are concerned, a complete specification of the relevant cold dark matter properties is given by the density parameter $\Omega_{\rm cdm}$, though those seeking to directly detect it are as interested in its interaction properties.

Cold dark matter is the standard assumption and gives an excellent fit to observations, except possibly on the shortest scales where there remains some controversy concerning the structure of dwarf galaxies and possible substructure in galaxy halos. It has long been excluded for all the dark matter to have a large velocity dispersion, so-called 'hot' dark matter, as it does not permit galaxies to form; for thermal relics the mass must be below about 1 keV to satisfy this constraint, though relics produced non-thermally, such as the axion, need not obey this limit. However, in future further parameters might need to be introduced to describe dark matter properties relevant to astrophysical observations. Suggestions which have been made include a modest velocity dispersion (warm dark matter) and dark matter self-interactions. There remains the possibility that the dark matter comprises two separate components, *e.g.*, a cold one and a hot one, an example being if massive neutrinos have a non-negligible effect.

**1.2.3.** *Dark energy*:

While the standard cosmological model given above features a cosmological constant, in order to explain observations indicating that the Universe is presently accelerating, further possibilities exist under the general heading 'dark energy'.[†] A particularly attractive possibility (usually called quintessence, though that word is used with various different meanings in the literature) is that a scalar field is responsible, with the mechanism mimicking that of early Universe inflation [15]. As described by Olive and Peacock, a fairly model-independent description of dark energy can be given just using the equation of state parameter $w$, with $w = -1$ corresponding to a cosmological constant. In general, the function $w$ could itself vary with redshift, though practical experiments devised so far would be sensitive primarily to some average value weighted over recent epochs. For high-precision predictions of CMB anisotropies, it is better to use a scalar-field description in order to have a self-consistent evolution of the 'sound speed' associated with the dark energy perturbations.

---

[†] Unfortunately this is rather a misnomer, as it is the negative pressure of this material, rather than its energy, that is responsible for giving the acceleration. Furthermore, while generally in physics matter and energy are interchangeable terms, dark matter and dark energy are quite distinct concepts.

A competing possibility is that the observed acceleration is due to a modification of gravity, *i.e.*, the left-hand side of Einstein's equation rather than the right. Observations of expansion kinematics alone cannot distinguish these two possibilities, but probes of the growth rate of structure formation may be able to.

Present observations are consistent with a cosmological constant, but often $w$ is kept as a free parameter to be added to the set described in the previous section. Most, but not all, researchers assume the weak energy condition $w \geq -1$. In the future, it may be necessary to use a more sophisticated parametrization of the dark energy.

### 1.2.4. *Complex ionization history*:

The full ionization history of the Universe is given by the ionization fraction as a function of redshift $z$. The simplest scenario takes the ionization to have the small residual value left after recombination up to some redshift $z_{\rm ion}$, at which point the Universe instantaneously reionizes completely. Then there is a one-to-one correspondence between $\tau$ and $z_{\rm ion}$ (that relation, however, also depending on other cosmological parameters). An accurate treatment of this process will track separate histories for hydrogen and helium. While currently rapid ionization appears to be a good approximation, as data improve a more complex ionization history may need to be considered.

### 1.2.5. *Varying 'constants'*:

Variation of the fundamental constants of Nature over cosmological times is another possible enhancement of the standard cosmology. There is a long history of study of variation of the gravitational constant $G$, and more recently attention has been drawn to the possibility of small fractional variations in the fine-structure constant. There is presently no observational evidence for the former, which is tightly constrained by a variety of measurements. Evidence for the latter has been claimed from studies of spectral line shifts in quasar spectra at redshifts of order two [16], but this is presently controversial and in need of further observational study.

More broadly, one can ask whether general relativity is valid at all epochs under consideration.

**1.2.6.**  *Cosmic topology*:

The usual hypothesis is that the Universe has the simplest topology consistent with its geometry, for example that a flat Universe extends forever. Observations cannot tell us whether that is true, but they can test the possibility of a non-trivial topology on scales up to roughly the present Hubble scale. Extra parameters would be needed to specify both the type and scale of the topology, for example, a cuboidal topology would need specification of the three principal axis lengths. At present, there is no direct evidence for cosmic topology, though the low values of the observed cosmic microwave quadrupole and octupole have been cited as a possible signature [17].

## 1.3. Probes

The goal of the observational cosmologist is to utilize astronomical information to derive cosmological parameters. The transformation from the observables to the key parameters usually involves many assumptions about the nature of the objects, as well as about the nature of the dark matter. Below we outline the physical processes involved in each probe, and the main recent results. The first two subsections concern probes of the homogeneous Universe, while the remainder consider constraints from perturbations.

In addition to statistical uncertainties we note three sources of systematic uncertainties that will apply to the cosmological parameters of interest: (i) due to the assumptions on the cosmological model and its priors (*i.e.*, the number of assumed cosmological parameters and their allowed range); (ii) due to the uncertainty in the astrophysics of the objects (*e.g.*, light curve fitting for supernovae or the mass–temperature relation of galaxy clusters); and (iii) due to instrumental and observational limitations (*e.g.*, the effect of 'seeing' on weak gravitational lensing measurements, or beam shape on CMB anisotropy measurements).

**1.3.1.**  *Direct measures of the Hubble constant*:

In 1929, Edwin Hubble discovered the law of expansion of the Universe by measuring distances to nearby galaxies. The slope of the relation between the distance and recession velocity is defined to be the Hubble constant $H_0$. Astronomers argued for decades on the systematic uncertainties in various methods and derived values over the wide range, $40\,\mathrm{km\,s^{-1}\,Mpc^{-1}} \lesssim H_0 \lesssim 100\,\mathrm{km\,s^{-1}\,Mpc^{-1}}$.

One of the most reliable results on the Hubble constant comes from the Hubble Space Telescope Key Project [18]. This study used the empirical period–luminosity relations for Cepheid variable stars to obtain distances to 31 galaxies, and calibrated a number of secondary distance indicators (Type Ia Supernovae, Tully–Fisher relation, surface brightness fluctuations, and Type II Supernovae) measured over distances of 400 to 600 Mpc. They estimated $H_0 = 72 \pm 3$ (statistical) $\pm 7$ (systematic) $\mathrm{km\,s^{-1}\,Mpc^{-1}}$.[‡] A recent study [19] of 240 Cepheids observed with an improved camera onboard the Hubble Space

---

[‡] Unless stated otherwise, all quoted uncertainties in this article are one-sigma/68% confidence. Cosmological parameters often have significantly non-Gaussian uncertainties. Throughout we have rounded central values, and especially uncertainties, from original sources in cases where they appear to be given to excessive precison.

Telescope has yielded an even more accurate figure, $H_0 = 74 \pm 4 \text{ km s}^{-1} \text{ Mpc}^{-1}$ (including both statistical and systematic errors). The major sources of uncertainty in these results are due to the heavy element abundance of the Cepheids and the distance to the fiducial nearby galaxy (called the Large Magellanic Cloud) relative to which all Cepheid distances are measured. Nevertheless, it is remarkable that this result is in such good agreement with the result derived from the WMAP CMB measurements combined with other probes (see Table 1.2).

**1.3.2.** *Supernovae as cosmological probes*:

The relation between observed flux and the intrinsic luminosity of an object depends on the luminosity distance $D_L$, which in turn depends on cosmological parameters. More specifically

$$D_L = (1+z) \, r_e(z) \,, \tag{1.10}$$

where $r_e(z)$ is the coordinate distance. For example, in a flat Universe

$$r_e(z) = \int_0^z \frac{dz'}{H(z')} \,. \tag{1.11}$$

For a general dark energy equation of state $w(z) = p_{\text{de}}(z)/\rho_{\text{de}}(z)$, the Hubble parameter is, still considering only the flat case,

$$\frac{H^2(z)}{H_0^2} = (1+z)^3 \, \Omega_m + \Omega_{\text{de}} \exp\left[3X(z)\right] \,, \tag{1.12}$$

where

$$X(z) = \int_0^z \left[1 + w(z')\right] \left(1 + z'\right)^{-1} dz' \,, \tag{1.13}$$

and $\Omega_m$ and $\Omega_{\text{de}}$ are the present density parameters of matter and dark energy components. If a general equation of state is allowed, then one has to solve for $w(z)$ (parametrized, for example, as $w(z) = w = \text{const.}$, or $w(z) = w_0 + w_1 z$) as well as for $\Omega_{\text{de}}$.

Empirically, the peak luminosity of supernovae of Type Ia (SNe Ia) can be used as an efficient distance indicator (*e.g.*, Ref. 20). The favorite theoretical explanation for SNe Ia is the thermonuclear disruption of carbon-oxygen white dwarfs. Although not perfect 'standard candles,' it has been demonstrated that by correcting for a relation between the light curve shape, color, and the luminosity at maximum brightness, the dispersion of the measured luminosities can be greatly reduced. There are several possible systematic effects which may affect the accuracy of the use of SNe Ia as distance indicators, for example, evolution with redshift and interstellar extinction in the host galaxy and in the Milky Way.

Two major studies, the 'Supernova Cosmology Project' and the 'High-$z$ Supernova Search Team', found evidence for an accelerating Universe [21], interpreted as due to a cosmological constant, or to a more general dark energy component. Current results from the 'Union sample' [22] of over 300 SNe Ia are shown in Fig. 1.1 (see also earlier results in Ref. 23). When combined with the CMB data (which indicates flatness, *i.e.*,

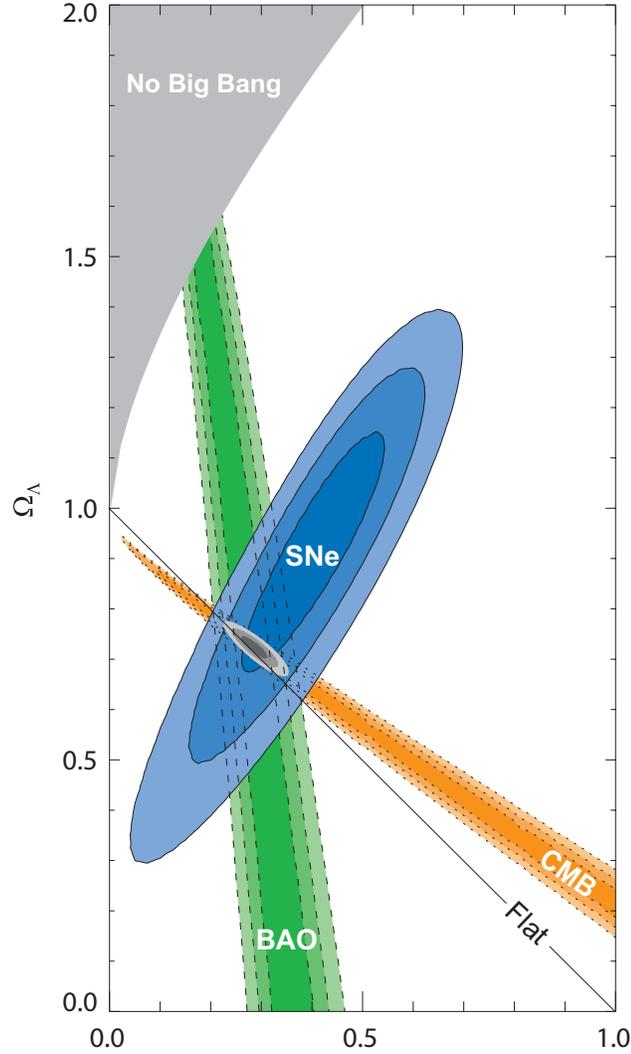

**Figure 1.1:** Confidence level contours of $68.3\%, 95.4\%$ and $99.7\%$ in the $\Omega_\Lambda$–$\Omega_{\rm m}$ plane from the Cosmic Microwave Background, Baryonic Acoustic Oscillations and the Union SNe Ia set, as well as their combination (assuming $w = -1$). [Courtesy of Kowalski *et al.* [22]]

$\Omega_{\rm m} + \Omega_\Lambda \approx 1$), the best-fit values are $\Omega_{\rm m} \approx 0.3$ and $\Omega_\Lambda \approx 0.7$. Most results in the literature are consistent with Einstein's $w = -1$ cosmological constant case.

For example, Kowalski *et al.* [22] deduced from SNe Ia combined with CMB and Baryonic Acoustic Oscillations (BAO) data (see next section), assuming a flat universe, that $w = -0.97 \pm +0.06 \pm 0.06$ (stat, sys) and $\Omega_{\rm m} = 0.274 \pm 0.016 \pm 0.012$. Similarly Kessler *et al.* [24] estimated $w = -0.96 \pm 0.06 \pm 0.12$ and $\Omega_{\rm m} = 0.265 \pm 0.016 \pm 0.025$, but they note a sensitivity to the light-curve fitter used.

Future experiments will aim to set constraints on the cosmic equation of state $w(z)$. However, given the integral relation between the luminosity distance and $w(z)$, it is not straightforward to recover $w(z)$ (*e.g.*, Ref. 25).

### 1.3.3. *Cosmic microwave background*:

The physics of the CMB is described in detail by Scott and Smoot in this volume. Before recombination, the baryons and photons are tightly coupled, and the perturbations oscillate in the potential wells generated primarily by the dark matter perturbations. After decoupling, the baryons are free to collapse into those potential wells. The CMB carries a record of conditions at the time of last scattering, often called primary anisotropies. In addition, it is affected by various processes as it propagates towards us, including the effect of a time-varying gravitational potential (the integrated Sachs-Wolfe effect), gravitational lensing, and scattering from ionized gas at low redshift.

The primary anisotropies, the integrated Sachs-Wolfe effect, and scattering from a homogeneous distribution of ionized gas, can all be calculated using linear perturbation theory, a widely-used implementation being the CMBFAST code of Seljak and Zaldarriaga [7] (CAMB is a popular alternative, often used embedded in the analysis package CosmoMC [26]). Gravitational lensing is also calculated in this code. Secondary effects such as inhomogeneities in the reionization process, and scattering from gravitationally-collapsed gas (the Sunyaev–Zel'dovich effect), require more complicated, and more uncertain, calculations.

The upshot is that the detailed pattern of anisotropies depends on all of the cosmological parameters. In a typical cosmology, the anisotropy power spectrum [usually plotted as $\ell(\ell+1)C_\ell$] features a flat plateau at large angular scales (small $\ell$), followed by a series of oscillatory features at higher angular scales, the first and most prominent being at around one degree ($\ell \simeq 200$). These features, known as acoustic peaks, represent the oscillations of the photon–baryon fluid around the time of decoupling. Some features can be closely related to specific parameters—for instance, the location of the first peak probes the spatial geometry, while the relative heights of the peaks probes the baryon density—but many other parameters combine to determine the overall shape.

The five-year data release from the WMAP satellite [1], henceforth WMAP5, has provided the most powerful results to date on the spectrum of CMB fluctuations, with a precision determination of the temperature power spectrum up to $\ell \simeq 900$, shown in Fig. 1.2, as well as measurements of the spectrum of $E$-polarization anisotropies and the correlation spectrum between temperature and polarization (those spectra having first been detected by DASI [27]). These are consistent with models based on the parameters we have described, and provide accurate determinations of many of those parameters [2].

WMAP5 provides an exquisite measurement of the location of the first acoustic peak, determining the angular-diameter distance of the last-scattering surface. In combination with other data this strongly constrains the spatial geometry, in a manner consistent with spatial flatness and excluding significantly-curved Universes. WMAP5 also gives a precision measurement of the age of the Universe. It gives a baryon density consistent with, and at higher precision than, that coming from BBN. It affirms the need for both dark matter and dark energy. It shows no evidence for dynamics of the dark energy, being consistent with a pure cosmological constant ($w = -1$). The density perturbations are consistent with a power-law primordial spectrum, with indications that the spectral slope may be less than the Harrison–Zel'dovich value $n = 1$ [2]. There is no indication of tensor perturbations, but the upper limit is quite weak. WMAP5's current best-fit for the reionization optical depth, $\tau = 0.087$, is in reasonable agreement with models of how early structure formation induces reionization.

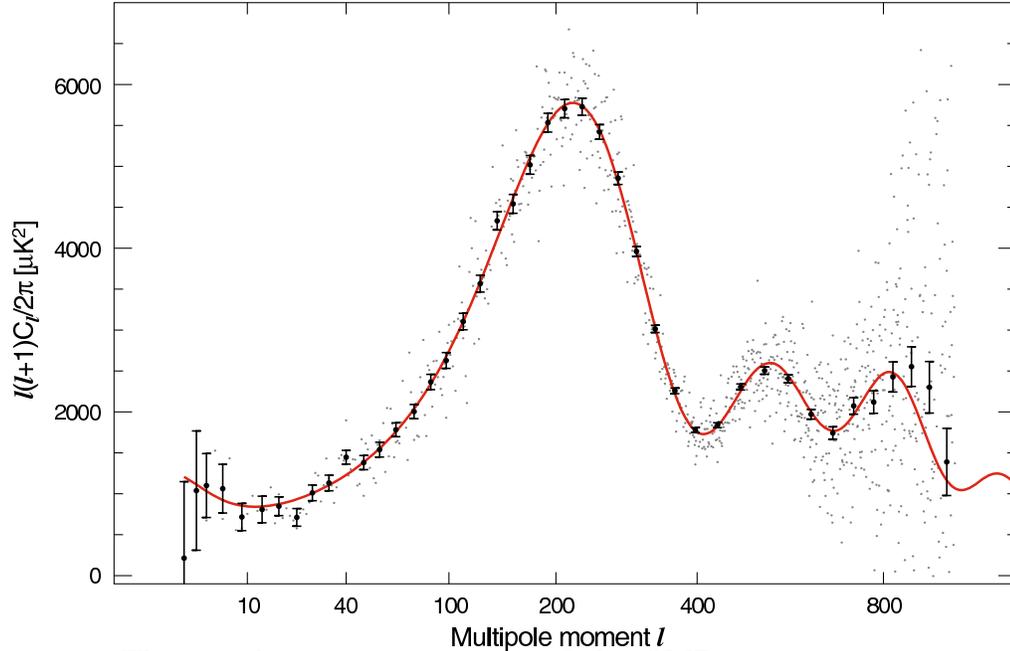

**Figure 1.2:** The angular power spectrum of the CMB temperature anisotropies from WMAP5 from Ref. [2]. The grey points are the unbinned data, and the solid are binned data with error estimates including cosmic variance. The solid line shows the prediction from the best-fitting $\Lambda$CDM model. [Figure courtesy NASA/WMAP Science Team.]

WMAP5 is consistent with other experiments and its dynamic range can be enhanced by including information from small-angle CMB experiments including ACBAR, CBI, and QUaD, which gives extra constraining power on some parameters.

**1.3.4.   *Galaxy clustering*:**

The power spectrum of density perturbations depends on the nature of the dark matter. Within the Cold Dark Matter model, the shape of the power spectrum depends primarily on the primordial power spectrum and on the combination $\Omega_{\rm m} h$, which determines the horizon scale at matter–radiation equality, with a subdominant dependence on the baryon density. The matter distribution is most easily probed by observing the galaxy distribution, but this must be done with care as the galaxies do not perfectly trace the dark matter distribution. Rather, they are a 'biased' tracer of the dark matter. The need to allow for such bias is emphasized by the observation that different types of galaxies show bias with respect to each other. Further, the observed 3D galaxy distribution is in redshift space, *i.e.*, the observed redshift is the sum of the Hubble expansion and the line-of-sight peculiar velocity, leading to linear and non-linear dynamical effects which also depend on the cosmological parameters. On the largest length scales, the galaxies are expected to trace the location of the dark matter, except for a constant multiplier $b$ to the power spectrum, known as the linear bias parameter. On scales smaller than 20 $h^{-1}$ Mpc or so, the clustering pattern is 'squashed' in the radial direction due to coherent infall, which depends approximately on the parameter $\beta \equiv \Omega_m^{0.6}/b$ (on these shorter scales, more complicated forms of biasing are not excluded by the data). On scales of a few $h^{-1}$ Mpc, there is an effect of elongation along the line of sight (colloquially known as the 'finger of God' effect) which depends on the galaxy velocity dispersion.

**1.3.4.1.** *Baryonic Acoustic Oscillations:*

The Fourier power spectra of the 2-degree Field (2dF) Galaxy Redshift Survey** and the Sloan Digital Sky Survey (SDSS)†† are well fitted by a ΛCDM model and both surveys show evidence for BAOs. Cole et al. [28] estimate from 2dF a baryon fraction $\Omega_b/\Omega_m = 0.18 \pm 0.05$ (1-$\sigma$ uncertainties). The shape of the power spectrum has been characterized by $\Omega_m h = 0.168 \pm 0.016$, and in combination with WMAP data gives $\Omega_m = 0.23 \pm 0.02$ (see also Ref. 29). Eisenstein *et al.* [30] reported a detection of the BAO peak in the large-scale correlation function of the SDSS sample of nearly 47,000 Luminous Red Galaxies (LRG). By using the baryon acoustic peak as a 'standard ruler' they found, independent of WMAP, that $\Omega_m = 0.27 \pm 0.03$ for a flat ΛCDM model. Signatures of BAOs have also been measured [31,32] from samples of nearly 600,000 LRGs with photometric redshifts (which are less accurate than spectroscopic redshifts, but easier to obtain for large samples).

The most recent work uses the SDSS LRG 7th Data Release [33,34]. Combining the so-called 'halo' power spectrum measurement with the WMAP5 results, for the flat ΛCDM model they find $\Omega_m = 0.289 \pm 0.019$ and $H_0 = 69.4 \pm 1.6$ km s$^{-1}$ Mpc$^{-1}$. Allowing for massive neutrinos in ΛCDM, they find $\sum m_\nu < 0.62$ eV at the 95% confidence level [33].

Combination of the 2dF data with the CMB indicates a 'biasing' parameter $b \sim 1$, in agreement with a 2dF-alone analysis of higher-order clustering statistics. However, results for biasing also depend on the length scale over which a fit is done, and the selection of the objects by luminosity, spectral type, or color. In particular, on scales smaller than 10 $h^{-1}$Mpc, different galaxy types are clustered differently. This 'biasing' introduces a systematic effect on the determination of cosmological parameters from redshift surveys. Prior knowledge from simulations of galaxy formation or from gravitational lensing data could help.

**1.3.4.2.** *Integrated Sachs–Wolfe effect:*

The integrated Sachs–Wolfe effect, described in the article by Scott and Smoot, is the change in CMB photon energy when propagating through the changing gravitational potential wells of developing cosmic structures. Correlating the large-angle CMB anisotropies with very large scale structures, first proposed in Ref. 35, has provided detections of this effect typically of significance 2 to $4\sigma$ [36]. As gravitational potentials do not evolve in critical density models, this provides direct evidence of a sub-critical matter density, and hence in combination with other probes supports the existence of dark energy.

---

** See http://www.mso.anu.edu.au/2dFGRS
†† See http://www.sdss.org

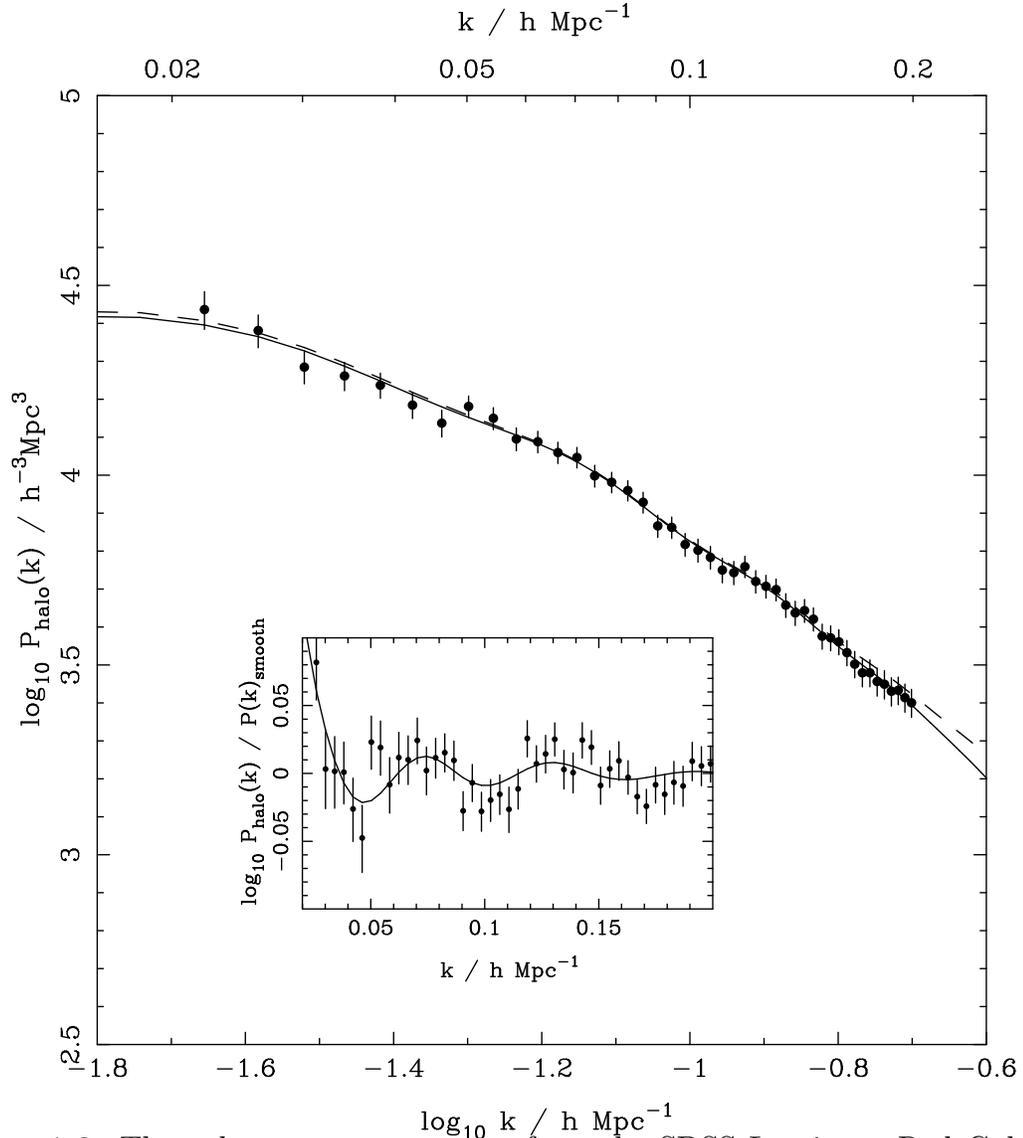

**Figure 1.3:** The galaxy power spectrum from the SDSS Luminous Red Galaxies (LRG). The best-fit LRG+WMAP $\Lambda$CDM model is shown for two sets of nuisance parameters (solid and dashed lines). The BAO inset shows the same data and model divided by a spline fit to the smooth component. [Figure provided by B. Reid and W. Percival; see Ref. 33 and Ref. 34.]

**1.3.4.3.** *Limits on neutrino mass from galaxy surveys and other probes:*

Large-scale structure data can put an upper limit on the ratio $\Omega_\nu/\Omega_m$ due to the neutrino 'free streaming' effect [37,38]. For example, by comparing the 2dF galaxy power spectrum with a four-component model (baryons, cold dark matter, a cosmological constant, and massive neutrinos), it is estimated that $\Omega_\nu/\Omega_m < 0.13$ (95% confidence limit), giving $\Omega_\nu < 0.04$ if a concordance prior of $\Omega_m = 0.3$ is imposed. The latter corresponds to an upper limit of about 2 eV on the total neutrino mass, assuming a prior of $h \approx 0.7$ [39]. Potential systematic effects include biasing of the galaxy distribution and non-linearities of the power spectrum. A similar upper limit of 2 eV has been derived from CMB anisotropies alone [40,41,42]. The above analyses assume that the primordial power spectrum is adiabatic, scale-invariant and Gaussian. Additional cosmological data sets have improved the results [43,44]. An upper limit on the total neutrino mass of 0.17 eV was reported by combining a large number of cosmological probes [45].

Laboratory limits on absolute neutrino masses from tritium beta decay and especially from neutrinoless double-beta decay should, within the next decade, push down towards (or perhaps even beyond) the 0.1 eV level that has cosmological significance.

**1.3.5.** *Clusters of galaxies*:

A cluster of galaxies is a large collection of galaxies held together by their mutual gravitational attraction. The largest ones are around $10^{15}$ Solar masses, and are the largest gravitationally-collapsed structures in the Universe. Even at the present epoch they are relatively rare, with only a few percent of galaxies being in clusters. They provide various ways to study the cosmological parameters; here we discuss constraints from the measurements of the cluster number density and the baryon fraction in clusters.

**1.3.5.1.** *Cluster number density:*

The first objects of a given kind form at the rare high peaks of the density distribution, and if the primordial density perturbations are Gaussian distributed, their number density is exponentially sensitive to the size of the perturbations, and hence can strongly constrain it. Clusters are an ideal application in the present Universe. They are usually used to constrain the amplitude $\sigma_8$, as a box of side $8\,h^{-1}\,\mathrm{Mpc}$ contains about the right amount of material to form a cluster. The most useful observations at present are of X-ray emission from hot gas lying within the cluster, whose temperature is typically a few keV, and which can be used to estimate the mass of the cluster. A theoretical prediction for the mass function of clusters can come either from semi-analytic arguments or from numerical simulations. At present, the main uncertainty is the relation between the observed gas temperature and the cluster mass, despite extensive study using simulations. Ref. [46] uses Chandra satellite X-ray data to obtain

$$\sigma_8 = 0.803 \pm 0.011 \text{ (stat.)} \pm 0.020 \text{ (sys.)} \qquad (68\% \text{ conf.}) \qquad (1.14)$$

for $\Omega_m = 0.25$. This result agrees well with the values predicted in cosmologies compatible with WMAP5.

The same approach can be adopted at high redshift (which for clusters means redshifts approaching one) to attempt to measure $\sigma_8$ at an earlier epoch. The evolution of $\sigma_8$ is primarily driven by the value of the matter density $\Omega_m$, with a sub-dominant dependence on the dark energy density. Such analyses favor a low matter density, again compatible with measurements from the CMB.

**1.3.5.2.**  *Cluster baryon fraction:*

If clusters are representative of the mass distribution in the Universe, the fraction of the mass in baryons to the overall mass distribution would be $f_\mathrm{b} = \Omega_\mathrm{b}/\Omega_\mathrm{m}$. If $\Omega_\mathrm{b}$, the baryon density parameter, can be inferred from the primordial nucleosynthesis abundance of the light elements, the cluster baryon fraction $f_\mathrm{b}$ can then be used to constrain $\Omega_\mathrm{m}$ and $h$ (*e.g.*, Ref. 47). The baryons in clusters are primarily in the form of X-ray-emitting gas that falls into the cluster, and secondarily in the form of stellar baryonic mass. Hence, the baryon fraction in clusters is estimated to be

$$f_\mathrm{b} = \frac{\Omega_\mathrm{b}}{\Omega_\mathrm{m}} \simeq f_\mathrm{gas} + f_\mathrm{gal}, \tag{1.15}$$

where $f_\mathrm{b} = M_\mathrm{b}/M_\mathrm{grav}$, $f_\mathrm{gas} = M_\mathrm{gas}/M_\mathrm{grav}$, $f_\mathrm{gal} = M_\mathrm{gal}/M_\mathrm{grav}$, and $M_\mathrm{grav}$ is the total gravitating mass. This leads to an approximate relation between $\Omega_\mathrm{m}$ and $h$:

$$\Omega_\mathrm{m} = \frac{\Omega_\mathrm{b}}{f_\mathrm{gas} + f_\mathrm{gal}} \simeq \frac{\Omega_\mathrm{b}}{0.08h^{-1.5} + 0.01h^{-1}}. \tag{1.16}$$

The ratio $\Omega_\mathrm{b}/\Omega_\mathrm{m}$ is consistent with other measures, and Allen *et al.* [48] give examples of constraints that can be obtained this way on both dark matter and dark energy using Chandra data across a range of redshifts.

**1.3.6.**  *Clustering in the inter-galactic medium*:

It is commonly assumed, based on hydrodynamic simulations, that the neutral hydrogen in the inter-galactic medium (IGM) can be related to the underlying mass distribution. It is then possible to estimate the matter power spectrum on scales of a few megaparsecs from the absorption observed in quasar spectra, the so-called Lyman-$\alpha$ forest. The usual procedure is to measure the power spectrum of the transmitted flux, and then to infer the mass power spectrum. Photo-ionization heating by the ultraviolet background radiation and adiabatic cooling by the expansion of the Universe combine to give a simple power-law relation between the gas temperature and the baryon density. It also follows that there is a power-law relation between the optical depth $\tau$ and $\rho_\mathrm{b}$. Therefore, the observed flux $F = \exp(-\tau)$ is strongly correlated with $\rho_\mathrm{b}$, which itself traces the mass density. The matter and flux power spectra can be related by

$$P_\mathrm{m}(k) = b^2(k)\ P_F(k), \tag{1.17}$$

where $b(k)$ is a bias function which is calibrated from simulations. Croft *et al.* [49] derived cosmological parameters from Keck Telescope observations of the Lyman-$\alpha$ forest at redshifts $z = 2 - 4$. Their derived power spectrum corresponds to that of a CDM model, which is in good agreement with the 2dF galaxy power spectrum. A recent study using VLT spectra [50] agrees with the flux power spectrum of Ref. 49. This method depends on various assumptions. Seljak *et al.* [51] pointed out that errors are sensitive to the range of cosmological parameters explored in the simulations, and the treatment of the mean transmitted flux. Nevertheless, this method has the potential of measuring accurately the power spectrum of mass fluctuations in a way different from the other methods.

**1.3.7.** *Gravitational lensing*:

Images of background galaxies get distorted due to the gravitational effect of mass fluctuations along the line of sight. Deep gravitational potential wells such as galaxy clusters generate 'strong lensing,' *i.e.*, arcs, arclets and multiple images, while more moderate fluctuations give rise to 'weak lensing'. Weak lensing is now widely used to measure the mass power spectrum in random regions of the sky (see Ref. 52 for recent reviews). As the signal is weak, the image of deformed galaxy shapes ('shear map') is analyzed statistically to measure the power spectrum, higher moments, and cosmological parameters.

The shear measurements are mainly sensitive to the combination of $\Omega_m$ and the amplitude $\sigma_8$. For example, the weak lensing signal detected by the CFHT Legacy Survey has been analyzed to yield $\sigma_8(\Omega_m/0.25)^{0.64} = 0.78 \pm 0.04$ [53] and $\sigma_8(\Omega_m/0.24)^{0.59} = 0.84 \pm 0.05$ [54] assuming a $\Lambda$CDM model. Earlier results are summarized in Ref. 52. There are various systematic effects in the interpretation of weak lensing, *e.g.*, due to atmospheric distortions during observations, the redshift distribution of the background galaxies, intrinsic correlation of galaxy shapes, and non-linear modeling uncertainties.

**1.3.8.** *Peculiar velocities*:

Deviations from the Hubble flow directly probe the mass fluctuations in the Universe, and hence provide a powerful probe of the dark matter. Peculiar velocities are deduced from the difference between the redshift and the distance of a galaxy. The observational difficulty is in accurately measuring distances to galaxies. Even the best distance indicators (*e.g.*, the Tully–Fisher relation) give an error of 15% per galaxy, hence limiting the application of the method at large distances. Peculiar velocities are mainly sensitive to $\Omega_m$, not to $\Omega_\Lambda$ or quintessence. Extensive analyses in the early 1990s (*e.g.*, Ref. 55) suggested a value of $\Omega_m$ close to unity. Further analysis [56], which takes into account non-linear corrections, gives $\sigma_8\Omega_m^{0.6} = 0.49 \pm 0.06$ and $\sigma_8\Omega_m^{0.6} = 0.63 \pm 0.08$ (90% errors) for two independent data sets. Analysis from pairwise velocities [57] gives $\sigma_8 = 1.1 \pm 0.2$, while bulk flows [58] give a lower limit $\sigma_8 > 1.11$ (95% CL), in disagreement with WMAP5. While at present cosmological parameters derived from peculiar velocities are strongly affected by random and systematic errors, a new generation of surveys may improve their accuracy. Three promising approaches are the 6dF near-infrared survey of 15,000 peculiar velocities[‡‡], supernovae Type Ia, and the kinematic Sunyaev–Zel'dovich effect.

There is also a renewed interest in 'redshift distortion'. As the measured redshift of a galaxy is the sum of its redshift due to the Hubble expansion and its peculiar velocity, this distortion depends on cosmological parameters (Ref. 59) via the growth rate $f(z) = d\ln\delta/d\ln a \approx \Omega^\gamma(z)$, where $\gamma = 0.55$ for a concordance $\Lambda$CDM model, and is different for a modified gravity model. Recent observational results [60] show that by measuring $f(z)$ with redshift it is feasible to constrain $\gamma$.

---

[‡‡] See http://www.mso.anu.edu.au/6dFGS/

**Table 1.2:** Parameter constraints reproduced from Dunkley *et al.* [2] and Komatsu *et al.* [3], with some additional rounding. All columns assume the ΛCDM cosmology with a power-law initial spectrum, no tensors, spatial flatness, and a cosmological constant as dark energy. Above the line are the six parameter combinations actually fit to the data; those below the line are derived from these. Two different data combinations are shown to highlight the extent to which this choice matters. The first column is WMAP5 alone, while the second column shows a combination of WMAP5 with BAO and SNe data as described in Ref. 3. The perturbation amplitude $\Delta_\mathcal{R}^2$ is specified at the scale $0.002\,\mathrm{Mpc}^{-1}$. Uncertainties are shown at 68% confidence, and caution is needed in extrapolating them to higher significance levels due to non-Gaussian likelihoods and assumed priors.

|  | WMAP5 alone | WMAP5 + BAO + SN |
|---|---|---|
| $\Omega_\mathrm{b} h^2$ | $0.0227 \pm 0.0006$ | $0.0227 \pm 0.0006$ |
| $\Omega_\mathrm{cdm} h^2$ | $0.110 \pm 0.006$ | $0.113 \pm 0.003$ |
| $\Omega_\Lambda$ | $0.74 \pm 0.03$ | $0.726 \pm 0.015$ |
| $n$ | $0.963^{+0.014}_{-0.015}$ | $0.960 \pm 0.013$ |
| $\tau$ | $0.087 \pm 0.017$ | $0.084 \pm 0.016$ |
| $\Delta_\mathcal{R}^2 \times 10^9$ | $2.41 \pm 0.11$ | $2.44 \pm 0.10$ |
| $h$ | $0.72 \pm 0.03$ | $0.705 \pm 0.013$ |
| $\sigma_8$ | $0.80 \pm 0.04$ | $0.81 \pm 0.03$ |
| $\Omega_\mathrm{m} h^2$ | $0.133 \pm 0.006$ | $0.136 \pm 0.004$ |

## 1.4. Bringing observations together

Although it contains two ingredients—dark matter and dark energy—which have not yet been verified by laboratory experiments, the ΛCDM model is almost universally accepted by cosmologists as the best description of the present data. The basic ingredients are given by the parameters listed in Sec. 1.1.4, with approximate values of some of the key parameters being $\Omega_\mathrm{b} \approx 0.04$, $\Omega_\mathrm{cdm} \approx 0.21$, $\Omega_\Lambda \approx 0.74$, and a Hubble constant $h \approx 0.72$. The spatial geometry is very close to flat (and usually assumed to be precisely flat), and the initial perturbations Gaussian, adiabatic, and nearly scale-invariant.

The most powerful single experiment is WMAP5, which on its own supports all these main tenets. Values for some parameters, as given in Dunkley *et al.* [2] and Komatsu *et al.* [3], are reproduced in Table 1.2. These particular results presume a flat Universe.

The constraints are somewhat strengthened by adding additional data-sets, as shown in the Table, though most of the constraining power resides in the WMAP5 data.

If the assumption of spatial flatness is lifted, it turns out that WMAP5 on its own only weakly constrains the spatial curvature, due to a parameter degeneracy in the angular-diameter distance. However inclusion of other data readily removes this, *e.g.*, inclusion of BAO and SNe data, plus the assumption that the dark energy is a cosmological constant, yields a constraint on $\Omega_{\rm tot} \equiv \sum \Omega_i + \Omega_\Lambda$ of [3]

$$\Omega_{\rm tot} = 1.006 \pm 0.006\,. \tag{1.18}$$

Results of this type are normally taken as justifying the restriction to flat cosmologies.

The baryon density $\Omega_{\rm b}$ is now measured with quite high accuracy from the CMB and large-scale structure, and is consistent with the determination from BBN; Fields and Sarkar in this volume quote the range $0.019 \leq \Omega_{\rm b} h^2 \leq 0.024$ (95% confidence).

While $\Omega_\Lambda$ is measured to be non-zero with very high confidence, there is no evidence of evolution of the dark energy density. The WMAP team find the limit $w < -0.86$ at 95% confidence from a compilation of data including SNe Ia, with the cosmological constant case $w = -1$ giving an excellent fit to the data.

The data provide strong support for the main predictions of the simplest inflation models: spatial flatness and adiabatic, Gaussian, nearly scale-invariant density perturbations. But it is disappointing that there is no sign of primordial gravitational waves, with WMAP5 alone providing only a weak upper limit $r < 0.43$ at 95% confidence [2] (this assumes no running, and weakens to 0.58 if running is allowed). The spectral index $n$ is placed in an interesting position by WMAP5, with indications that $n < 1$ is required by the data. However, the confidence with which $n = 1$ is ruled out is still rather weak, and in our view it is premature to conclude that $n = 1$ is no longer viable.

Tests have been made for various types of non-Gaussianity, a particular example being a parameter $f_{\rm NL}$ which measures a quadratic contribution to the perturbations. Tests distinguish between non-Gaussianity of 'local' and 'equilateral' type (see Ref. [3] for details), and current constraints give $-9 < f_{\rm NL}^{\rm local} < 110$ and $-150 < f_{\rm NL}^{\rm equil} < 250$ at 95% confidence (these look weak, but prominent non-Gaussianity requires the product $f_{\rm NL} \Delta_{\mathcal{R}}$ to be large, and $\Delta_{\mathcal{R}}$ is of order $10^{-5}$). It will be interesting to watch if the tendency of the former to have a positive value reaches significance in future data.

One parameter which is very robust is the age of the Universe, as there is a useful coincidence that for a flat Universe the position of the first peak is strongly correlated with the age. The WMAP5 result is $13.69 \pm 0.13$ Gyr (assuming flatness). This is in good agreement with the ages of the oldest globular clusters [61] and radioactive dating [62].

## 1.5.   Outlook for the future

The concordance model is now well established, and there seems little room left for any dramatic revision of this paradigm. A measure of the strength of that statement is how difficult it has proven to formulate convincing alternatives.

Should there indeed be no major revision of the current paradigm, we can expect future developments to take one of two directions. Either the existing parameter set will continue to prove sufficient to explain the data, with the parameters subject to ever-tightening constraints, or it will become necessary to deploy new parameters. The latter outcome would be very much the more interesting, offering a route towards understanding new physical processes relevant to the cosmological evolution. There are many possibilities on offer for striking discoveries, for example:

- The cosmological effects of a neutrino mass may be unambiguously detected, shedding light on fundamental neutrino properties;
- Compelling detection of deviations from scale-invariance in the initial perturbations would indicate dynamical processes during perturbation generation by, for instance, inflation;
- Detection of primordial non-Gaussianities would indicate that non-linear processes influence the perturbation generation mechanism;
- Detection of variation in the dark energy density (*i.e.*, $w \neq -1$) would provide much-needed experimental input into the question of the properties of the dark energy.

These provide more than enough motivation for continued efforts to test the cosmological model and improve its precision.

Over the coming years, there are a wide range of new observations which will bring further precision to cosmological studies. Indeed, there are far too many for us to be able to mention them all here, and so we will just highlight a few areas.

The CMB observations will improve in several directions. The current frontier is the study of polarization, first detected in 2002 by DASI and for which power spectrum measurements have now been made by several experiments. Future measurements may be able to separately detect the two modes of polarization. Another area of development is pushing accurate power spectrum measurements to smaller angular scales, with the Atacama Cosmology Telescope (ACT) and South Pole Telescope (SPT) both now in operation. Finally, we mention the Planck satellite, launched in May 2009, which will make high-precision all-sky maps of temperature and polarization, utilizing a very wide frequency range for observations to improve understanding of foreground contaminants, and to compile a large sample of clusters via the Sunyaev–Zel'dovich effect.

On the supernova side, the most ambitious initiatives at present are satellite missions JDEM (Joint Dark Energy Mission), proposed to NASA and DOE, and Euclid proposed to ESA. An impressive array of ground-based dark energy surveys are also already operational, under construction, or proposed, including the ESSENCE project, the Dark Energy Survey, Pan-Starrs, and LSST. With large samples, it may be possible to detect evolution of the dark energy density, thus measuring its equation of state and perhaps even its variation.

An exciting new area for the future will be radio surveys of the redshifted 21-cm line of hydrogen. Because of the intrinsic narrowness of this line, by tuning of the bandpass the emission from narrow redshift slices of the Universe will be measured to extremely high redshift, probing the details of the reionization process at redshifts up to perhaps 20. LOFAR is the first instrument able to do this and is at an advanced construction stage. In the longer term, the Square Kilometer Array (SKA) will take these studies to a precision level.

The above future surveys will address fundamental questions of physics well beyond just testing the 'concordance' $\Lambda$CDM model and minor variations. By learning about both the geometry of the universe and the growth of perturbations, it would be possible to test theories of modified gravity and inhomogeneous universes.

The development of the first precision cosmological model is a major achievement. However, it is important not to lose sight of the motivation for developing such a model, which is to understand the underlying physical processes at work governing the Universe's evolution. On that side, progress has been much less dramatic. For instance, there are many proposals for the nature of the dark matter, but no consensus as to which is correct. The nature of the dark energy remains a mystery. Even the baryon density, now measured to an accuracy of a few percent, lacks an underlying theory able to predict it even within orders of magnitude. Precision cosmology may have arrived, but at present many key questions remain unanswered.